\documentclass[11pt]{article}

\usepackage[top=1in,bottom=1in,left=1in,right=1in]{geometry}
\usepackage{natbib}
\usepackage[unicode=true,bookmarks=true,bookmarksnumbered=false,bookmarksopen=false,breaklinks=false,pdfborder={0 0 1},backref=false,colorlinks=true]{hyperref}
\hypersetup{citecolor=blue}
\usepackage{url}

\usepackage{amsmath}
\usepackage{mathrsfs}
\usepackage{amssymb}
\usepackage{amsthm}
\usepackage[normalem]{ulem}

\newtheorem{theorem}{Theorem}

\newtheorem{corollary}{Corollary}

\usepackage{float}
\usepackage{rotfloat}

\floatstyle{ruled}
\newfloat{algorithm}{tpb}{loa}
\providecommand{\algorithmname}{Algorithm}
\floatname{algorithm}{\protect\algorithmname}

\usepackage{algpseudocode,algorithm}
\algnewcommand\algorithmicinput{\textbf{Input}:}
\algnewcommand\algorithmicoutput{\textbf{Output}:}
\algnewcommand\INPUT{\item[\algorithmicinput]}
\algnewcommand\OUTPUT{\item[\algorithmicoutput]}

\usepackage{graphicx}
\usepackage[caption=false,labelformat=simple]{subfig}

\usepackage{array}
\newcolumntype{L}[1]{>{\raggedright\let\newline\\\arraybackslash\hspace{0pt}}m{#1}}
\newcolumntype{C}[1]{>{\centering\let\newline\\\arraybackslash\hspace{0pt}}m{#1}}
\newcolumntype{R}[1]{>{\raggedleft\let\newline\\\arraybackslash\hspace{0pt}}m{#1}}

\usepackage{rotating}
\usepackage{multirow}

\usepackage{tikz}
\usetikzlibrary{shapes,arrows,backgrounds,calc,positioning,fit,petri,plotmarks}
\usetikzlibrary{arrows}

\usepackage{enumerate}
\usepackage{paralist}

\newcommand*{\affaddr}[1]{#1} 
\newcommand*{\affmark}[1][*]{\textsuperscript{#1}}

\global\long\def\bX{\mathbf{X}}
\global\long\def\bx{\mathbf{x}}
\global\long\def\bY{\mathbf{Y}}

\global\long\def\bW{\mathbf{W}}
\global\long\def\bw{\mathbf{w}}
\global\long\def\bU{\mathbf{U}}

\global\long\def\bV{\mathbf{V}}

\global\long\def\bP{\mathbf{P}}
\global\long\def\bQ{\mathbf{Q}}
\global\long\def\bR{\mathbf{R}}
\global\long\def\bM{\mathbf{M}}
\global\long\def\bm{\mathbf{m}}
\global\long\def\bS{\mathbf{S}}

\global\long\def\bvarepsilon{\boldsymbol{\varepsilon}}
\global\long\def\bSigma{\boldsymbol{\Sigma}}

\global\long\def\bdeta{\boldsymbol{\eta}}
\global\long\def\btheta{\boldsymbol{\theta}}
\global\long\def\bphi{\boldsymbol{\phi}}
\global\long\def\bPhi{\boldsymbol{\Phi}}
\global\long\def\bpsi{\boldsymbol{\psi}}
\global\long\def\bPsi{\boldsymbol{\Psi}}
\global\long\def\bLambda{\boldsymbol{\Lambda}}
\global\long\def\bDelta{\boldsymbol{\Delta}}
\global\long\def\bXi{\boldsymbol{\Xi}}
\global\long\def\bPi{\boldsymbol{\Pi}}

\usepackage{xr}
\makeatletter
\newcommand*{\addFileDependency}[1]{
  \typeout{(#1)}
  \@addtofilelist{#1}
  \IfFileExists{#1}{}{\typeout{No file #1.}}
}
\makeatother

\title{Mediation Analysis with Multiple Exposures and Multiple Mediators}

\author{%
    Yi Zhao\affmark[1] and for the Alzheimer's Disease Neuroimaging
Initiative\footnote{Data used in preparation of this article were obtained from the Alzheimer's Disease Neuroimaging Initiative (ADNI) database (\url{adni.loni.usc.edu}). As such, the investigators within the ADNI contributed to the design and implementation of ADNI and/or provided data but did not participate in analysis or writing of this report. A complete list of ADNI investigators can be found at: \url{http://adni.loni.usc.edu/wp-content/uploads/how_to_apply/ADNI_Acknowledgement_List.pdf}} \\
    \affaddr{\affmark[1]Department of Biostatistics and Health Data Science, Indiana University School of Medicine} \\
}

\date{}

\providecommand{\keywords}[1]
{
  {\small   
  \textbf{Keywords:} #1 }
}

\begin{document}

\maketitle

\thispagestyle{empty}

\begin{abstract}
    A mediation analysis approach is proposed for multiple exposures, multiple mediators, and a continuous scalar outcome under the linear structural equation modeling framework. It assumes that there exist orthogonal components that demonstrate parallel mediation mechanisms on the outcome, and thus is named \textbf{P}rincipal \textbf{C}omponent \textbf{M}ediation \textbf{A}nalysis (PCMA). Likelihood-based estimators are introduced for simultaneous estimation of the component projections and effect parameters. The asymptotic distribution of the estimators is derived for low-dimensional data. A bootstrap procedure is introduced for inference. Simulation studies illustrate the superior performance of the proposed approach. Applied to a proteomics-imaging dataset from the Alzheimer's Disease Neuroimaging Initiative (ADNI), the proposed framework identifies protein deposition -- brain atrophy -- memory deficit mechanisms consistent with existing knowledge and suggests potential AD pathology by integrating data collected from different modalities.
\end{abstract}



\clearpage
\setcounter{page}{1}

\section{Introduction}
\label{sec:intro}

In this study, a mediation analysis framework for cases with multiple exposures, multiple mediators, and a continuous scalar outcome is introduced. It assumes that there exist orthogonal components that constitute parallel mediation mechanisms on the outcome, and thus is named \textbf{P}rincipal \textbf{C}omponent \textbf{M}ediation \textbf{A}nalysis (PCMA). Linear structural equation models are proposed for such parallel mechanisms and likelihood-based estimators are proposed for simultaneous identification of the component projections and estimation of the model coefficients. 

The proposed framework is motivated by a proteomic-imaging study from the Alzheimer's Disease Neuroimaging Initiative (ADNI). AD is an irreversible neurodegenerative disease. In 2021, an estimate of $6.5$ million American aged $65$ and older are living with AD and this number is projected to grow to $15.0$ million by 2060~\citep{wiley2021alzheimer}. As there exists no effective treatment for AD, understanding the disease pathology, identifying treatment targets, and developing early diagnosis and intervention strategies are critically important. The ADNI was launched in 2003 aiming to identify AD biomarkers and measure the disease progression through multiple modalities, including magnetic resonance imaging (MRI), positron emission tomography (PET), other biological specimens, and clinical and neuropsychological assessments. In this study, we focus on two modalities, the cerebrospinal fluid (CSF) proteomics, and T1-weighted MR brain images, and investigate the connections between modalities, as well as to neural cognitive behaviors. 
It is widely known that amyloid-$\beta$ (A$\beta$) and tau are two key protein markers for AD, where A$\beta$ is the main component of amyloid plaques and tau is the main component of neurofibrillary tangles. In AD pathophysiological models, it is hypothesized that A$\beta$ and tau deposition can disrupt cell-to-cell communications and destroy brain cells, which leads to structural atrophy in areas such as the medial temporal lobe and ultimately deficits in cognitive functions~\citep{mormino2009episodic}. The CSF contains proteins related to various biological processes. With high-throughput proteomics technology, many proteins have been consistently identified whose alterations are associated with AD. Among these, many have also been verified to be related to A$\beta$ and/or tau pathology~\citep{wesenhagen2020cerebrospinal}.
A stereotypical pattern of neurodegeneration suggests that brain atrophy occurs early in the medial temporal lobe and soon spreads to the rest of the cortical areas following the  temporal--parietal--frontal trajectory, while motor areas are not generally impacted until later stages of AD~\citep{pini2016brain}. Based on the pattern of intraneuronal neurofibrillary changes, \citet{braak1991neuropathological} characterized six stages and marked affected brain regions under each stage. For example, the entorhinal cortex (located in the medial temporal lobe) is a Braak I region affected by tau propagation first. 
In this study, we aim to expand the study scope by considering a data-driven approach to identify mechanisms of CSF protein alteration--brain atrophy--memory decline using CSF proteomics, brain volumetric, and memory behavior measurement data from ADNI.

Based on the biological assumptions, we formulate the problem as a mediation analysis with multiple exposures (intensity of CSF proteins/peptides), multiple mediators (volume of brain regions), and a continuous scalar outcome (measurement of memory behavior). Mediation analysis aims to delineate the underlying mechanism between the exposure and the outcome by decomposing the exposure effect into the part through the mediator (called the indirect effect) and the part not through the mediator (called the direct effect). Cases with a single exposure and a single mediator have been extensively studied~\cite[see a review by ][and references therein]{vanderweele2016mediation} and various extensions for different data types, such as time-to-event data~\citep{tchetgen2011causal}, time series~\citep{zhao2019granger}, and functional data~\citep{lindquist2012functional,zeng2021causal}, have been introduced. With the spread utilization of data acquired from high-throughput technologies, approaches for mediation analysis are adapted to dealing with high-dimensional mediators, where high-dimensional omics or neuroimaging data are considered as the mediator candidates aiming to identify significant biological mediation pathways~\cite[examples include][among many others]{chen2017high,song2018bayesian,zhao2022pathway}. Approaches for both multiple exposures and multiple mediators are relatively sparse. In a recent work, \citet{aung2020application} proposed to reduce the dimension of the exposures into a small number of groups based on prior domain knowledge and verified that the inter-group correlation is negligible. Though \citet{long2020framework} considered gene expression and protein measures as the exposures and mediators, respectively, the proposed mediation framework implicitly assumed no interference between the exposures and applied the principal component analysis (PCA) on the mRNA data first. In the study, only five metabolic proteins were preselected as potential mediators. \citet{zhang2021high} developed two regularization procedures for a large number of exposures and mediators, but the procedures require the mediators to be independent, which is an overly stringent assumption for brain volumetric data. Recently, \citet{zhao2022multimodal} introduced a framework for high-dimensional exposures and high-dimensional mediators, which combines PCA with regularized mediation estimation for mediator selection. The PCA is applied to the exposures to create independent components, which can be considered a data processing step. One drawback of this approach is without consider the connections to the mediators and outcome when doing the PCA. In addition, with regularized estimators, the post-selection inference is not straightforward.

In mediation analysis, one popular parametric approach is the linear structural equation modeling (LSEM) framework. One example is the \citet{baron1986moderator} approach, where two regression models are assumed, one for the mediator and one for the outcome. For a regression problem with multivariate dependent and multivariate independent variables, partial least squares (PLS) and its extensions are widely implemented. In a PLS regression, latent structures of the dependent and independent variables that demonstrate the strongest associations are identified. In this study, we adopt this idea to mediation analysis to identify orthogonal latent structures of the exposures and mediators that constitute parallel mediation mechanisms. With multiple exposures and mediators, when the mediation paths are parallel, it greatly simplifies the problem as it can be handled separately by performing a series of marginal mediation analyses~\citep{imai2013identification,vanderweele2015explanation}. For a mediation analysis, the LSEM framework has two regression models. Thus, different from a PLS regression, information contained in both regression models should be taken into consideration when identifying the latent structures and estimating the model parameters. As the exposures, mediators, and outcome are all assumed to be continuous variables in this study, by imposing normality assumptions, the joint likelihood function of the LSEMs is derived and estimators that maximize the joint likelihood function are introduced to identify the latent structures and effect parameters.

The rest of the manuscript is organized as the following. Section~\ref{sec:model} introduces a mediation analysis approach for the scenario of multiple exposures and multiple mediators under the LSEM framework, where the multivariate data are projected into lower-dimensional spaces for mediation mechanisms. Likelihood-based estimators are introduced and an estimation algorithm is proposed for simultaneous identification of the projecting vectors and model coefficients. Asymptotic distribution of the estimators are derived under the lower-dimensional case. A bootstrap procedure is introduced for inference. In Section~\ref{sec:sim}, simulation studies are presented to demonstrate the performance of the proposal. Section~\ref{sec:ADNI} applies the mediation framework to the ADNI proteomic-imaging study. A discussion on the identified proteins, brain regions, and mediation paths are included. Section~\ref{sec:discussion} summarizes the manuscript with discussions.




\section{Model and Method}
\label{sec:model}

For subject $i\in\{1,\dots,n\}$, where $n$ is the number of subjects, let $\bx_{i}=(x_{i1},\dots,x_{ip})^\top\in\mathbb{R}^{p}$ denote the $p$-dimensional exposures, $\bm_{i}=(m_{i1},\dots,m_{iq})^\top\in\mathbb{R}^{q}$ denote the $q$-dimensional mediators, and $y_{i}\in\mathbb{R}$ denote the scalar outcome. Let $\bw_{i}=(w_{i1},\dots,w_{is})^\top\in\mathbb{R}^{s}$ denote the $s$-dimensional covariates (with the first element of one for the intercept). These can be pre-exposure covariates and/or post-exposure mediator-outcome confounding factors not induced by the exposures. 
In the ADNI application, $\bx_{i}$ is the intensity of $p=35$ CSF proteins, $\bm_{i}$ is the brain volumetric data extracted from $q=37$ regions of interest using an atlas by \citet{doshi2016muse}, $y_{i}$ is a composite memory score (ADNI\_MEM) calculated from a battery of neuropsychological tests, and $\bw_{i}$ includes age, gender, and years of education.
Let $\bX=(\bx_{1},\dots,\bx_{n})^\top\in\mathbb{R}^{n\times p}$, $\bM=(\bm_{1},\dots,\bm_{n})^\top\in\mathbb{R}^{n\times q}$, $\bW=(\bw_{1},\dots,\bw_{n})^\top\in\mathbb{R}^{n\times s}$, and $\bY=(y_{1},\dots,y_{n})^\top\in\mathbb{R}^{n}$. It is assumed that there exist linear projections, $\bphi\in\mathbb{R}^{p}$ and $\bpsi\in\mathbb{R}^{q}$, such that after projection, the variables satisfy the following linear structural equation models (LSEMs).
\begin{eqnarray}
    \bM\bpsi &=& \bX\bphi\cdot\alpha+\bW\btheta_{1}+\bvarepsilon, \label{eq:model_M} \\
    \bY &=& \bX\bphi\cdot\gamma+\bM\bpsi\cdot\beta+\bW\btheta_{2}+\bdeta, \label{eq:model_Y}
\end{eqnarray}
where $\alpha,\beta,\gamma$, $\btheta_{1}\in\mathbb{R}^{s},\btheta_{2}\in\mathbb{R}^{s}$ are model coefficients, $\bvarepsilon=(\varepsilon_{1},\dots,\varepsilon_{n})^\top\in\mathbb{R}^{n}$ and $\bdeta=(\eta_{1},\dots,\eta_{n})^\top\in\mathbb{R}^{n}$ are model errors. $\bvarepsilon$ is assumed to be independent of $\bX$ and $\bW$, and $\bdeta$ is independent of $\bX$, $\bM$, and $\bW$. Under the assumption that there exists no unmeasured mediator-outcome confounder, $\bvarepsilon$ is independent of $\bdeta$. For continuous outcome and mediator, it is assumed that the errors follow normal distributions, where $\epsilon_{i}$ follows a normal distribution with mean zero and variance $\sigma^{2}$ and $\eta_{i}$ follows a normal distribution with mean zero and variance $\tau^{2}$. 

\begin{figure}
    \begin{center}
        \includegraphics[width=0.6\textwidth,page=1]{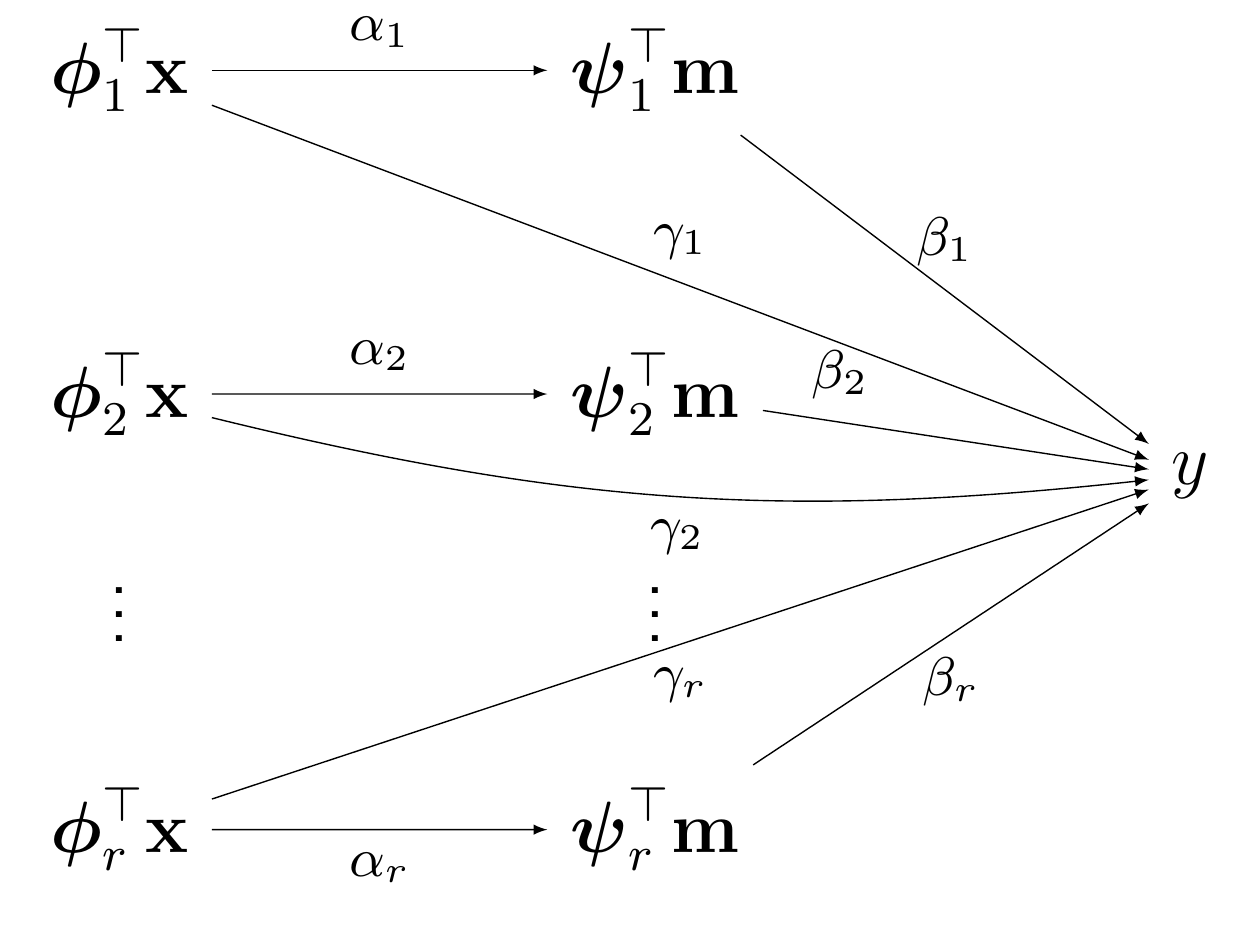}
    \end{center}
    \caption{\label{fig:DAG}A conceptual diagram with $r$ parallel mediation paths. $\bphi_{1}^\top\bx,\dots,\bphi_{r}^\top\bx$ are $r$ orthogonal exposure components. $\bpsi_{j}^\top\bm$ is the mediator component associated with $\bphi_{j}^\top\bx$, for $j=1,\dots,r$.}
\end{figure}

Under Models~\eqref{eq:model_M} and~\eqref{eq:model_Y}, it is assumed that for a linear projection of the exposures, a linear projection of the mediators captures the mediation effect on the outcome. $\bphi^\top\bx$ is called an exposure component and $\bpsi^\top\bm$ is the associated mediator component.
In the ADNI application, CSF protein intensities are the exposures and brain volumetric data are the mediators. With proper thresholding or sparsifying, the projections identify a combination of proteins and brain regions enabling a network-level interpretation of the mediation mechanism.
When there exist multiple projections of the exposures, they are assumed to be orthogonal to each other. In other words, there is no interference among the exposure components. Under this assumption, the path effects are parallel to each other, and one can fit LSEMs separately. In Section~\ref{sub:estimation}, an algorithm will be introduced to identify multiple components sequentially. Figure~\ref{fig:DAG} presents a conceptual mechanistic diagram with $r<\min(p,q)$ parallel mediation paths. In the figure, the confounding factors ($\bw$) are removed to focus on the mediation mechanisms. Under the proposed models, $\mathrm{DE}(\bphi_{j}^\top\bx)=\gamma_{j}$ is the direct effect of the $j$th exposure component on the outcome, $\mathrm{IE}(\bphi_{j}^\top\bx,\bpsi_{j}^\top\bm)=\alpha_{j}\beta_{j}$ is the indirect effect of the $j$th exposure component on the outcome through the corresponding mediator component, and $\mathrm{TE}(\bphi_{j}^\top\bx)=\mathrm{DE}(\bphi_{j}^\top\bx)+\mathrm{IE}(\bphi_{j}^\top\bx,\bpsi_{j}^\top\bm)=\gamma_{j}+\alpha_{j}\beta_{j}$ is the total effect of the $j$th exposure component.

\subsection{Estimation}
\label{sub:estimation}

In Models~\eqref{eq:model_M} and~\eqref{eq:model_Y}, not only the model coefficients but also the projections need to be estimated. Let $\Theta=(\bphi,\bpsi,\alpha,\beta,\gamma,\btheta_{1},\btheta_{2},\sigma^{2},\tau^{2})$ denote the parameter set. Under the normality assumption, it is proposed to estimate the parameters by minimizing the negative joint likelihood function given by the following (after removing the constant terms)
\begin{equation}
    \ell(\Theta)=\frac{1}{\sigma^{2}}\|\bM\bpsi-\bX\bphi\alpha-\bW\btheta_{1}\|_{2}^{2}+\frac{1}{\tau^{2}}\|\bY-\bX\bphi\gamma-\bM\bpsi\beta-\bW\btheta_{2}\|_{2}^{2}+n\log\sigma^{2}+n\log\tau^{2}.
\end{equation}
Note that the parameters are not identifiable due to the existence of the products, $\bX\bphi\alpha$, $\bX\bphi\gamma$, and $\bM\bpsi\beta$. To uniquely estimate the parameters, it is imposed that $\bphi$ and $\bpsi$ both have the vector norm of one. The objective function is then written as
\begin{eqnarray}\label{eq:obj_likelihood}
    \text{minimize} && \ell(\Theta), \nonumber \\
    \text{such that} && \|\bphi\|_{2}=1, \quad \|\bpsi\|_{2}=1.
\end{eqnarray}
It is easy to show that $\ell$ is bi-convex. Thus, we propose to estimate the parameters by (block) coordinate descent. In the following, we discuss the solution to each parameter. For $\bphi$, assuming $(\bpsi,\alpha,\beta,\gamma,\btheta_{1},\btheta_{2},\sigma^{2},\tau^{2})$ is given, because of the constraint, the Lagrangian form is
\begin{equation}
    \mathcal{L}(\bphi,\lambda_{1})=\ell(\bphi)+\lambda_{1}(\bphi^\top\bphi-1),
\end{equation}
where $\lambda_{1}$ is the Lagrangian parameter. By taking partial derivatives and setting to zero,
\begin{equation}\label{eq:solution_phi}
    \hat{\bphi}=\left\{\left(\frac{\alpha^{2}}{\sigma^{2}}+\frac{\gamma^{2}}{\tau^{2}}\right)\bX^\top\bX+\lambda_{1}\boldsymbol{\mathrm{I}}\right\}^{-1}\underbrace{\bX^\top\left\{\left(\frac{\alpha}{\sigma^{2}}-\frac{\beta\gamma}{\tau^{2}}\right)\bM\bpsi+\frac{\gamma}{\tau^{2}}\bY-\left(\frac{\alpha}{\sigma^{2}}\bW\btheta_{1}+\frac{\gamma}{\tau^{2}}\bW\btheta_{2}\right)\right\}}_{\bU},
\end{equation}
and $\lambda_{1}$ is the solution to
\begin{equation}\label{eq:solution_lambda1}
    \bphi^\top\bphi-1=\bU^\top\left\{\left(\frac{\alpha^{2}}{\sigma^{2}}+\frac{\gamma^{2}}{\tau^{2}}\right)\bX^\top\bX+\lambda_{1}\boldsymbol{\mathrm{I}}\right\}^{-2}\bU-1=0,
\end{equation}
which can be solved by any algorithm that finds the unique root of a function. Analogously for $\bpsi$ (given $\bphi,\alpha,\beta,\gamma,\btheta_{1},\btheta_{2},\sigma^{2},\tau^{2}$), the solution is
\begin{equation}\label{eq:solution_psi}
    \hat{\bpsi}=\left\{\left(\frac{1}{\sigma^{2}}+\frac{\beta^{2}}{\tau^{2}}\right)\bM^\top\bM+\lambda_{2}\boldsymbol{\mathrm{I}}\right\}^{-1}\underbrace{\bM^\top\left\{\left(\frac{\alpha}{\sigma^{2}}-\frac{\beta\gamma}{\tau^{2}}\right)\bX\bphi+\frac{\beta}{\tau^{2}}\bY+\left(\frac{1}{\sigma^{2}}\bW\btheta_{1}-\frac{\beta}{\tau^{2}}\bW\btheta_{2}\right)\right\}}_{\bV},
\end{equation}
where $\lambda_{2}$ is the Lagrangian parameter and the value is the solution to
\begin{equation}\label{eq:solution_lambda2}
    \bpsi^\top\bpsi-1=\bV^\top\left\{\left(\frac{1}{\sigma^{2}}+\frac{\beta^{2}}{\tau^{2}}\right)\bM^\top\bM+\lambda_{2}\boldsymbol{\mathrm{I}}\right\}^{-2}\bV-1=0.
\end{equation}
For the model coefficients,
\begin{eqnarray}\label{eq:solution_coef}
    \hat{\alpha} &=& (\bphi^\top\bX^\top\bX\bphi)^{-1}(\bphi^\top\bX^\top)(\bM\bpsi+\bW\btheta_{1}), \nonumber \\
    \hat{\beta} &=& (\bpsi^\top\bM^\top\bM\bpsi)^{-1}(\bpsi^\top\bM^\top)(\bY-\gamma\bX\bphi-\bW\btheta_{2}), \nonumber \\
    \hat{\gamma} &=& (\bphi^\top\bX^\top\bX\bphi)^{-1}(\bphi^\top\bX^\top)(\bY-\beta\bW\bpsi-\bW\btheta_{2}), \nonumber \\
    \hat{\btheta}_{1} &=& (\bW^\top\bW)^{-1}\bW^\top(\bM\bpsi-\alpha\bX\bphi), \nonumber \\
    \hat{\btheta}_{2} &=& (\bW^\top\bW)^{-1}\bW^\top(\bY-\gamma\bX\bphi-\beta\bM\bpsi).
\end{eqnarray}
For the variances,
\begin{eqnarray}\label{eq:solution_var}
    \hat{\sigma}^{2} &=& \frac{1}{n}\|\bM\bpsi-\bX\bphi\alpha-\bW\btheta_{1} \|_{2}^{2}, \nonumber \\
    \hat{\tau}^{2} &=& \frac{1}{n}\|\bY-\bX\bphi\gamma-\bM\bpsi\beta-\bW\btheta_{2}\|_{2}^{2}.
\end{eqnarray}
Algorithm~\ref{alg:opt_likelihood} summarizes the estimation procedure to simultaneously identify the first exposure and mediator component and estimate the effect parameters. In practice, occasions, such as $\bphi$ or $\bpsi$ or both are prespecified, may occur. For example, the $p$ exposures are mutually independent. For the $j$th exposure, one can specify $\bphi_{j}$ as a $p$-dimensional vector with the $j$th element one and the rest zero. Under these occasions, one can modify Algorithm~\ref{alg:opt_likelihood} by setting the corresponding projection vector(s) to the prespecified value and optimizing over the rest parameters.

\begin{algorithm}[!ph]
    \caption{\label{alg:opt_likelihood}The optimization algorithm for \eqref{eq:obj_likelihood}.}
    \begin{algorithmic}[1]
    \INPUT $(\bX,\bM,\bW,\bY)$.
    \State \textbf{initialization}: $\left\{\bphi^{(0)},\bpsi^{(0)},\alpha^{(0)},\beta^{(0)},\gamma^{(0)},\btheta_{1}^{(0)},\btheta_{2}^{(0)}, \sigma^{2(0)},\tau^{2(0)}\right\}$.
    \Repeat    
    \State update $\bphi^{(s+1)}$ given $\bpsi^{(s)},\alpha^{(s)}, \beta^{(s)}, \gamma^{(s)}, \btheta_{1}^{(s)},\btheta_{2}^{(s)},\sigma^{2(s)},\tau^{2(s)}$ by \eqref{eq:solution_phi} and \eqref{eq:solution_lambda1}.
    \State update $\bpsi^{(s+1)}$ given $\bphi^{(s+1)},\alpha^{(s)}, \beta^{(s)}, \gamma^{(s)}, \btheta_{1}^{(s)},\btheta_{2}^{(s)},\sigma^{2(s)},\tau^{2(s)}$ by \eqref{eq:solution_psi} and \eqref{eq:solution_lambda2}.
    \State update $\alpha^{(s+1)},\beta^{(s+1)},\gamma^{(s+1)},\btheta_{1}^{(s+1)},\btheta_{2}^{(s+1)}$ given $\bphi^{(s+1)},\bpsi^{(s+1)},\sigma^{2(s)},\tau^{2(s)}$ by \eqref{eq:solution_coef}.
    \State update $\sigma^{2(s+1)},\tau^{2(s+1)}$ given $\bphi^{(s+1)},\bpsi^{(s+1)},\alpha^{(s+1)}, \beta^{(s+1)}, \gamma^{(s+1)}, \btheta_{1}^{(s+1)},\btheta_{2}^{(s+1)}$ by \eqref{eq:solution_var}.
    \Until{the stopping criterion is met.}
    \OUTPUT $\left\{\hat{\bphi},\hat{\bpsi},\hat{\alpha},\hat{\beta},\hat{\gamma},\hat{\btheta}_{1},\hat{\btheta}_{2},\hat{\sigma}^{2},\hat{\tau}^{2}\right\}$.
    \end{algorithmic}
\end{algorithm}

To identify higher-order components, as the orthogonality assumptions are imposed, it is proposed to remove the identified components from the data, similar to the identification of higher-order components in the principal component analysis. Let $\hat{\bPhi}^{(k)}=(\hat{\bphi}_{1},\dots,\hat{\bphi}_{k})\in\mathbb{R}^{p\times k}$ and $\hat{\bPsi}^{(k)}=(\hat{\bpsi}_{1},\dots,\hat{\bpsi}_{k})\in\mathbb{R}^{q\times k}$ denote the first $k$ projections of the exposures and mediators, respectively. For $i=1,\dots,n$, let
\begin{equation}
    \hat{\bx}_{i}^{(k+1)}=\bx_{i}-\bx_{i}\hat{\bPhi}^{(k)}\hat{\bPhi}^{(k)\top} \quad \text{and} \quad \hat{\bm}_{i}^{(k+1)}=\bm_{i}-\bm_{i}\hat{\bPsi}^{(k)}\hat{\bPsi}^{(k)\top}
\end{equation}
be the new exposure and mediator, respectively. As the parallel assumption is imposed on the mediation paths, the effects from the identified components should be removed from the outcome making the next mediation mechanism orthogonal to the existing ones. Let $\{\hat{\gamma}_{j}\}_{j=1}^{k}$ and $\{\hat{\beta}_{j}\}_{j=1}^{k}$ denote the estimated model coefficients in~\eqref{eq:model_Y} of the first $k$ components. For $i=1,\dots,n$, let
\begin{equation}
    \hat{y}_{i}^{(k+1)}=y_{i}-\sum_{j=1}^{k}(\hat{\bphi}_{j}^\top\bx_{i})\hat{\gamma}_{j}-\sum_{j=1}^{k}(\hat{\bpsi}_{j}^\top\bm_{i})\hat{\beta}_{j}
\end{equation}
be the new outcome. Applying Algorithm~\ref{alg:opt_likelihood} to the new data, $\{\hat{\bx}_{i}^{(k+1)},\hat{\bm}_{i}^{(k+1)},\hat{y}_{i}^{(k+1)},\bw_{i}\}$, one can identify the $(k+1)$th component of $\bphi$ and $\bpsi$, which are orthogonal to $\hat{\bPhi}^{(k)}$ and $\hat{\bPsi}^{(k)}$, respectively. By performing a data manipulation rather than imposing orthogonality constraints, the computational efficiency is significantly improved, especially when the dimensions are large. To determine the number of components, we propose to choose the number based on the significance of the indirect effect. In Section~\ref{sub:inference}, a bootstrap approach is proposed to draw inference on the direct and indirect effects. As in most cases, the study interest is in the indirect effect, the number of components can be then chosen as the first $r$ components with a significant indirect effect.

\subsection{Asymptotic properties}
\label{sub:asmp}

In this section, the asymptotic properties of the proposed estimators are discussed under the scenario that $p,q\ll n$ and $p,q$ are fixed. As normality is assumed and the proposed estimators are maximum likelihood estimators, likelihood-based asymptotic theories can be applied. The following theorem, Theorem~\ref{thm:asmp}, gives the asymptotic distribution of the proposed estimators.
\begin{theorem}\label{thm:asmp}
    For $p,q\ll n$ and fixed, assume that as $n\rightarrow \infty$, $\bX^\top\bX/n\rightarrow\bP\in\mathbb{R}^{p\times p}$, $\bM^\top\bM/n\rightarrow\bQ\in\mathbb{R}^{q\times q}$, $\bX^\top\bM/n\rightarrow\bR\in\mathbb{R}^{p\times q}$, $\bphi^\top\bX^\top\bX\bphi/n\rightarrow\kappa_{x}$, $\bpsi^\top\bM^\top\bM\bpsi/n\rightarrow\kappa_{m}$, $\bphi^\top\bX^\top\bM\bpsi/n\rightarrow\kappa_{xm}$, and $\bW^\top\bW/n\rightarrow \bS\in\mathbb{R}^{s\times s}$. Under Models~\eqref{eq:model_M} and~\eqref{eq:model_Y} and normally distributed errors, the estimators proposed in Section~\ref{sub:estimation} have the following asymptotic distributions.
    \begin{equation}
        \sqrt{n}\left(\begin{pmatrix}
         \hat{\bphi} \\
         \hat{\bpsi}
        \end{pmatrix}-\begin{pmatrix}
         \bphi \\
         \bpsi
    \end{pmatrix}\right)\overset{\mathcal{D}}{\longrightarrow}\mathcal{N}\left(\boldsymbol{\mathrm{0}},\bPi\right),
    \end{equation}
    \[
        \text{where } \bPi^{-1}=\begin{pmatrix}
     (\alpha^{2}/\sigma^{2}+\gamma^{2}/\tau^{2})\bP & -(\alpha/\sigma^{2}-\beta\gamma/\tau^{2})\bR \\
     -(\alpha/\sigma^{2}-\beta\gamma/\tau^{2})\bR^\top & (1/\sigma^{2}+\beta^{2}/\tau^{2})\bQ
    \end{pmatrix};
    \]
    \begin{equation}
        \sqrt{n}\left(\begin{pmatrix}
         \hat{\alpha} \\
         \hat{\beta} \\
         \hat{\gamma}
        \end{pmatrix}-\begin{pmatrix}
         \alpha \\
         \beta \\
         \gamma
        \end{pmatrix}\right)\overset{\mathcal{D}}{\longrightarrow}\mathcal{N}\left(\boldsymbol{\mathrm{0}},\bXi\right), \quad \text{where } \bXi^{-1}=\begin{pmatrix}
     \kappa_{x}/\sigma^{2} & 0 & 0 \\
     0 & \kappa_{m}/\tau^{2} & \kappa_{xm}/\tau^{2} \\
     0 & \kappa_{xm}/\tau^{2} & \kappa_{x}/\tau^{2}
    \end{pmatrix};
    \end{equation}
    \begin{equation}
        \sqrt{n}\left(\begin{pmatrix}
            \hat{\btheta}_{1} \\
            \hat{\btheta}_{2}
        \end{pmatrix}-\begin{pmatrix}
            \btheta_{1} \\
            \btheta_{2}
        \end{pmatrix}\right)\overset{\mathcal{D}}{\longrightarrow}\mathcal{N}\left(\boldsymbol{\mathrm{0}},\begin{pmatrix}
            \sigma^{2}\bS^{-1} & \boldsymbol{\mathrm{0}} \\
            \boldsymbol{\mathrm{0}} & \tau^{2}\bS^{-1}
        \end{pmatrix}\right).
    \end{equation}
\end{theorem}
\begin{corollary}\label{corol:asmp_AB}
    Assume the assumptions in Theorem~\ref{thm:asmp} hold, the estimator of the indirect effect, $\hat{\alpha}\hat{\beta}$ has the following asymptotic distribution,
    \begin{equation}
        \sqrt{n}\left(\hat{\alpha}\hat{\beta}-\alpha\beta\right)\overset{\mathcal{D}}{\longrightarrow}\mathcal{N}\left(0,\sigma_{\alpha\beta}^{2}\right), \quad \text{where } \sigma_{\alpha\beta}^{2}=\frac{\beta^{2}\sigma^{2}}{\kappa_{x}}+\frac{\alpha^{2}\tau^{2}\kappa_{x}}{\kappa_{x}\kappa_{m}-\kappa_{xm}^{2}}.
    \end{equation}
\end{corollary}

\subsection{Inference}
\label{sub:inference}

Theorem~\ref{thm:asmp} offers the asymptotic distribution of the proposed estimators for the parameters in the LSEMs. For the indirect effect, denoted as the product of $\alpha$ and $\beta$, the asymptotic distribution can be derived from the Delta method (Corollary~\ref{corol:asmp_AB}). However, considering the fact that in finite sample, the distribution of $\hat{\alpha}_{j}\hat{\beta}_{j}$ can be far from Gaussian, a nonparametric bootstrap procedure is proposed to perform inference on the direct and indirect effects. The following gives the steps for the $j$th component.
\begin{description}
    \item[Step 1.] Generate a bootstrap sample $\{(\hat{\bphi}_{j}^\top\bx_{i})^{*},(\hat{\bpsi}_{j}^\top\bm_{i})^{*},\bw_{i}^{*},y_{i}^{*}\}$ of size $n$ by sampling with replacement.
    \item[Step 2.] Estimate the model coefficients and variances using Algorithm~\ref{alg:opt_likelihood} with $(\hat{\bphi}_{j},\hat{\bpsi}_{j})$ known.
    \item[Step 3.] Repeat Steps 1--2 for $B$ times.
    \item[Step 4.] Construct bootstrap confidence intervals for the direct and indirect effects using either the percentile or bias-corrected approach~\citep{efron1987better} under a pre-specified significance level.
\end{description}



\section{Simulation Study}
\label{sec:sim}

We first use simulation studies to examine the performance of the proposed framework, named PCMA (\textbf{P}rincipal \textbf{C}omponent \textbf{M}ediation \textbf{A}nalysis). As existing approaches for mediation analysis with multiple exposures and multiple mediators are scarce, an approach derived from the PCA-based high-dimensional mediation analysis introduced by \citet{huang2016hypothesis} is considered and named PCA-HP. This approach includes two steps: (i) conduct PCA on the exposures and treat top PCs accounting for at least $85\%$ of the data variation as the independent exposures; (ii) for each exposure component, apply the approach in \citet{huang2016hypothesis} and test for significant mediator components. 

In the study, $r=2$ parallel mediation paths are considered to be significant.
The exposures, $\bx$, are generated from a multivariate normal distribution with mean zero and covariance matrix $\bSigma_{x}$. $\bSigma_{x}$ has the eigendecomposition of $\bSigma_{x}=\bPhi\bLambda\bPhi^\top$, where $\bPhi\in\mathbb{R}^{p\times p}$ is an orthonormal matrix and $\bLambda\in\mathbb{R}^{p\times p}$ is a diagonal matrix with eigenvalues exponentially decay. 
The mediators, $\bm$, are assumed to follow a multivariate normal distribution with covariance matrix $\bSigma_{m}$, where $\bSigma_{m}=\bPsi\bDelta\bPsi^\top$ with $\bSigma\in\mathbb{R}^{q\times q}$ an orthonormal matrix and $\bDelta\in\mathbb{R}^{q\times q}$ a diagonal matrix. To generate $\bm$, the PCs of $\bm$ (denoted as $\tilde{\bm}$) are generated first and then transform back as $\bm=\bPhi\tilde{\bm}$. The first two PCs are generated following model~\eqref{eq:model_M}, where $\varepsilon$'s are from the standard normal distribution, and the rest are normally distributed with mean zero and variance as the corresponding eigenvalue. The outcome, $y$, is generated following model~\eqref{eq:model_Y} by adding up the two mediation paths. The model error, $\eta$, is from the standard normal distribution. For the two mediation paths, $(\alpha_{1},\beta_{1},\gamma_{1})=(2,2,1)$ and $(\alpha_{2},\beta_{2},\gamma_{2})=(2,1,-1)$, thus $(\mathrm{IE}_{1},\mathrm{DE}_{1})=(4,1)$ and $(\mathrm{IE}_{2},\mathrm{DE}_{2})=(2,-1)$. In this simulation study, for demonstration purposes, covariates are not considered. In practice, one can either include covariates in the models or treat covariate adjustment as a data processing step to remove the confounding effects~\citep{rosenbaum2002covariance}. Two scenarios of data dimension are considered: (1) $p=5$, $q=10$ and (2) $p=35$, $q=37$ (the same as the ADNI dataset in Section~\ref{sec:ADNI}). 
To evaluate the performance, the following metrics are considered. For the projection vectors, the magnitude of the inner products, denoted as $|\langle \hat{\bphi},\bphi\rangle|$ and $|\langle \hat{\bpsi},\bpsi \rangle|$, is introduced as a similarity metric between the estimate and truth. A frequency of successfully identifying the components is reported, where success is defined when the average of the two magnitudes is greater than $0.5$. For the coefficients ($\alpha,\beta,\gamma$) and indirect effect (IE), estimation bias, standard error (SE), and mean squared error (MSE) are reported. Simulations are repeated $200$ times.

Table~\ref{table:sim_est} presents the simulation results. For both scenarios, the proposed PCMA approach correctly identifies the two components $100\%$ of the time. The PCA-HP approach only identifies the first component (C1) in scenario (1) and the percentage of identifying the target components is less than $100\%$ in most of the cases. For scenario (1), the similarity of $\bphi$ and $\bpsi$ estimates are both high over $0.900$ using the PCMA approach. While the similarity of $\bpsi$ estimate using the PCA-HP approach is around $0.600$ and an increase in the sample size does not improve the performance. When the data dimension increases to $p=35$ and $q=37$, the similarity of $\bphi$ and $\bpsi$ estimates from the PCMA approach is lower at the sample size of $n=135$. When the sample size increases to $n=500$, the performance of the PCMA approach significantly improves with similarities over $0.900$. When comparing the SE of the similarities, the PCMA approach is much lower suggesting a more stable performance in identifying the components. For both scenarios, when comparing the performance in estimating the model parameters, the performance of the PCMA approach is superior with a lower bias, SE, and MSE.

\begin{sidewaystable}
    \caption{\label{table:sim_est}Simulation results from $200$ replicates. $\%$: percentage of identifying the corresponding component; DE: direct effect; IE: indirect effect; SE: standard error; MSE: mean squared error; C1/C2: the two components with a significant mediation effect.}
    \begin{center}
        \resizebox{\textwidth}{!}{
        \begin{tabular}{c c l c r c c r r r c r r r c r r r c r r r}
            \hline
            & & & & & & & \multicolumn{3}{c}{$\alpha$} && \multicolumn{3}{c}{$\beta$} && \multicolumn{3}{c}{$\gamma$ ($\mathrm{DE}$)} && \multicolumn{3}{c}{$\mathrm{IE}$} \\
            \cline{8-10}\cline{12-14}\cline{16-18}\cline{20-22}
            \multicolumn{1}{c}{\multirow{-2}{*}{$(p,q)$}} & \multicolumn{1}{c}{\multirow{-2}{*}{$n$}} & \multicolumn{1}{c}{\multirow{-2}{*}{Method}} & \multicolumn{1}{c}{\multirow{-2}{*}{}} & \multicolumn{1}{c}{\multirow{-2}{*}{$\%$}} & \multicolumn{1}{c}{\multirow{-2}{*}{$|
            \langle \hat{\bphi},\bphi \rangle|$ (SE)}} & \multicolumn{1}{c}{\multirow{-2}{*}{$|
            \langle \hat{\bpsi},\bpsi \rangle|$ (SE)}} & \multicolumn{1}{c}{Bias} & \multicolumn{1}{c}{SE} & \multicolumn{1}{c}{MSE} && \multicolumn{1}{c}{Bias} & \multicolumn{1}{c}{SE} & \multicolumn{1}{c}{MSE} && \multicolumn{1}{c}{Bias} & \multicolumn{1}{c}{SE} & \multicolumn{1}{c}{MSE} && \multicolumn{1}{c}{Bias} & \multicolumn{1}{c}{SE} & \multicolumn{1}{c}{MSE} \\
            \hline
            & & & C1 & $57\%$ & $0.998~(0.002)$ & $0.633~(0.096)$ & $-0.736$ & $0.195$ & $0.579$ && $-0.742$ & $0.218$ & $0.597$ && $2.357$ & $0.540$ & $5.846$ && $-2.372$ & $0.540$ & $5.918$ \\
            & & \multirow{-2}{*}{PCA-HP} & C2 & $0\%$ & - & - & - & - & - && - & - & - && - & - & - && - & - & - \\
            \cline{3-22}
            & & & C1 & $100\%$ & $0.930~(0.055)$ & $0.969~(0.005)$ & $0.093$ & $0.142$ & $0.029$ && $0.212$ & $0.137$ & $0.063$ && $-0.228$ & $0.292$ & $0.137$ && $0.629$ & $0.408$ & $0.562$ \\
            & \multirow{-4}{*}{$100$} & \multirow{-2}{*}{PCMA} & C2 & $100\%$ & $0.946~(0.041)$ & $0.943~(0.017)$ & $0.017$ & $0.124$ & $0.016$ && $-0.411$ & $0.111$ & $0.181$ && $-0.230$ & $0.231$ & $0.106$ && $-0.815$ & $0.222$ & $0.713$ \\
            \cline{2-22}
            & & & C1 & $68\%$ & $0.999~(0.001)$ & $0.622~(0.085)$ & $-0.756$ & $0.172$ & $0.601$ && $-0.752$ & $0.171$ & $0.594$ && $2.414$ & $0.447$ & $6.026$ && $-2.420$ & $0.448$ & $6.053$ \\
            & & \multirow{-2}{*}{PCA-HP} & C2 & $0\%$ & - & - & - & - & - && - & - & - && - & - & - && - & - & - \\
            \cline{3-22}
            & & & C1 & $100\%$ & $0.967~(0.014)$ & $0.976~(0.001)$ & $0.020$ & $0.033$ & $0.001$ && $0.169$& $0.053$ & $0.031$ && $-0.212$ & $0.108$ & $0.057$ && $0.381$ & $0.135$ & $0.163$ \\
            \multirow{-8}{*}{$(5,10)$} & \multirow{-4}{*}{$500$} & \multirow{-2}{*}{PCMA} & C2 & $100\%$ & $0.969~(0.015)$ & $0.971~(0.003)$ & $0.012$ & $0.045$ & $0.002$ && $-0.425$ & $0.048$ & $0.183$ && $-0.202$ & $0.099$ & $0.051$ && $-0.844$ & $0.095$ & $0.722$ \\
            \hline
            & & & C1 & $96\%$ & $0.899~(0.082)$ & $0.855~(0.104)$ & $-0.307$ & $0.322$ & $0.198$ && $-0.005$ & $0.353$ & $0.124$ && $0.039$ & $0.876$ & $0.765$ && $-0.589$ & $0.967$ & $1.278$ \\
            & & \multirow{-2}{*}{PCA-HP} & C2 & $100\%$ & $0.843~(0.098)$ & $0.904~(0.092)$ & $-0.313$ & $0.293$ & $0.183$ && $-0.423$ & $0.778$ & $0.781$ && $0.952$ & $0.588$ & $1.251$ && $-1.003$ & $1.301$ & $2.691$ \\
            \cline{3-22}
            & & & C1 & $100\%$ & $0.785~(0.068)$ & $0.893~(0.022)$ & $0.295$ & $0.219$ & $0.135$ && $0.487$ & $0.207$ & $0.280$ && $-0.300$ & $0.441$ & $0.283$ && $1.711$ & $0.741$ & $3.474$ \\
            & \multirow{-4}{*}{$135$} & \multirow{-2}{*}{PCMA} & C2 & $100\%$ & $0.751~(0.073)$ & $0.695~(0.055)$ & $-0.127$ & $0.216$ & $0.062$ && $0.093$ & $0.239$ & $0.065$ && $-1.057$ & $0.486$ & $.1353$ && $0.042$ & $0.482$ & $0.233$ \\
            \cline{2-22}
            & & & C1 & $90\%$ & $0.968~(0.033)$ & $0.878~(0.102)$ & $-0.230$ & $0.192$ & $0.089$ && $-0.039$ & $0.304$ & $0.093$ && $0.337$ & $0.824$ & $0.789$ && $-0.479$ & $0.822$ & $0.901$ \\
            & & \multirow{-2}{*}{PCA-HP} & C2 & $99\%$ & $0.943~(0.046)$ & $0.943~(0.074)$ & $-0.149$ & $0.129$ & $0.039$ && $-0.378$ & $0.453$ & $0.348$ && $0.773$ & $0.237$ & $0.654$ && $-0.850$ & $0.851$ & $1.443$ \\
            \cline{3-22}
            & & & C1 & $100\%$ & $0.920~(0.024)$ & $0.960~(0.005)$ & $0.089$ & $0.058$ & $0.011$ && $0.219$ & $0.065$ & $0.052$ && $-0.198$ & $0.129$ & $0.056$ && $0.637$ & $0.181$ & $0.438$ \\
            \multirow{-8}{*}{$(35,37)$} & \multirow{-4}{*}{$500$} & \multirow{-2}{*}{PCMA} & C2 & $100\%$ & $0.930~(0.019)$ & $0.911~(0.018)$ & $-0.038$ & $0.059$ & $0.005$ && $-0.336$ & $0.056$ & $0.116$ && $-0.315$ & $0.111$ & $0.112$ && $-0.699$ & $0.110$ & $0.501$ \\
            \hline
        \end{tabular}
        }
    \end{center}
\end{sidewaystable}




\section{The Proteomics-Imaging Study of AD}
\label{sec:ADNI}


In this section, the proposed approach is applied to a proteomic-imaging dataset from the ADNI study. The CSF proteomics data were acquired from targeted liquid chromatography multiple reaction monitoring mass spectrometry (LC/MS-MRM). MRM is a highly specific, sensitive, and reproducible label-free technique for targeted protein quantification. The objective of the original study is to examine the ability of a penal of MS-measured peptides in discriminating disease status. The list was selected based on their detection record in CSF and relevance to AD. These compounds were sent to the detector and went through a series of processing procedures, including peak integration, outliers detection, normalization, quantification, and quality control using test/retest samples. 
The intensity of a list of $p=35$ proteins is considered as the exposures ($\bX$) based on existing findings of AD proteomics studies~\citep{wesenhagen2020cerebrospinal}. A brain imaging measure obtained from anatomical magnetic resonance imaging (MRI) is considered as the mediator ($\bM$). After following standard data preprocessing steps, volumetric measures from $q=37$ regions of interest spanning the whole brain~\citep{doshi2016muse} were extracted. To adjust for variations in individual brain size, the volume of each brain region is standardized by the total intracranial volume. A cognitive outcome, called ADNI-MEM, is considered as the outcome variable ($Y$). It is a composite memory score calculated from a battery of neuropsychological tests. Covariates considered include age, gender, and years of education ($\bW$). In this study, data from $n=135$ subjects diagnosed with mild cognitive impairment (MCI) at recruitment are analyzed. As a prodromal stage of AD, subjects with MCI experience a slight but noticeable and measurable cognitive decline and an increased risk of developing AD or other types of dementia. Thus, understanding the disease pathology of MCI plays a key role in guiding early diagnosis and intervention for AD.

The proposed approach identifies three orthogonal components with a significant indirect effect, denoted as C1, C2, and C3. Table~\ref{table:adni_coef} presents the estimates and $95\%$ confidence intervals from $1000$ bootstrap samples. The indirect effect of all three components are significant and negative, while the underlying mechanism varies. In C1, $\alpha$ is significantly positive and $\beta$ is significantly negative; in C2 and C3, $\alpha$ is significantly negative and $\beta$ is significantly positive. For C1, as the intensity of the protein component increases, the volume of the corresponding brain component becomes relatively larger, while leading to a decrease in the memory score. For C2 and C3, it is the opposite that as the protein intensity increases, the brain volume decreases, leading to a memory decline.
In order to better interpret the exposure and mediator components, an \textit{ad hoc} procedure is employed to sparsify the loading profile using the lasso regularization~\citep{tibshirani1996regression}, similar to the sparse principal component analysis approach~\citep{zou2006sparse}. 
Table~\ref{table:adni_protein} lists the top-loaded proteins in the components and Table~\ref{tabel:adni_brain} presents the top-loaded brain regions.

In Table~\ref{table:adni_protein}, the direction of protein level in MCI/AD compared to normal control reported in the existing literature is summarized. From the table, the sign of the loadings are in line with the (consistent) directions in C1 and mostly consistent in C2 and C3. For example, in C1, proteins with a positive loading are downregulated in MCI/AD and proteins with a negative loading are upregulated. 
Here, we pick a few proteins and discuss their roles in AD pathology. 
NPTX1 and NPTX2 are from the family of long neuronal pentraxins. The family has the function of binding AMPA type glutamate receptors and contributes to synaptic plasticity during neurodevelopment and adulthood. Reduction in NPTX2 together with amyloidosis was found to induce a synergistic reduction in inhibitory circuit function. It was also found to be related to hippocampal volume and cognitive decline among AD patients~\citep{xiao2017nptx2}. 
NRXN1 is a transmembrane protein found in presynaptic terminals. The function of such protein is to form pairs with postsynaptic neuroligins facilitating neuronal connections. CSF concentration of NRXN1 was found downregulated in AD patients~\citep{brinkmalm2018parallel}.
VGF is a neuropeptide precursor. It was found to potentially serve a protective role against AD as an over-expression of VGF rescued A$\beta$-mediated memory impairment~\citep{beckmann2020multiscale}.
PTGDS is one of the most abundant proteins in the CSF. Existing findings suggest it to be an endogenous A$\beta$ chaperone and thus plays an important role in AD pathology~\citep{kanekiyo2007lipocalin}.
Focusing on the AD group, a significantly positive association between CH3L1 and A$\beta$, as well as CH3L1 and tau, was identified~\citep{heywood2015identification,dayon2018alzheimer}. Compared to controls, CSF CH3L1 level was found to be significantly elevated in AD patients~\citep{heywood2015identification,paterson2016targeted}.
Human KLK6 is primarily abundant in the spinal cord, brain stem, hippocampus, and thalamus and has been found to be relevant in both A$\beta$ and tau pathology~\citep{angelo2006substrate,goldhardt2019kallikrein}.
ENPP2 has also been found an AD marker serving a contributory role in AD pathology including A$\beta$ formation, increased tau-phosphorylation, and neurite retraction in neuronal cells. An elevated level in the CSF and a differential expression in the frontal cortex were observed when compared to controls~\citep{umemura2006autotaxin,heywood2015identification}.

\begin{figure}
    \begin{center}
        \subfloat[\label{fig:adni_brain_map_C1}C1]{\includegraphics[width=0.33\textwidth]{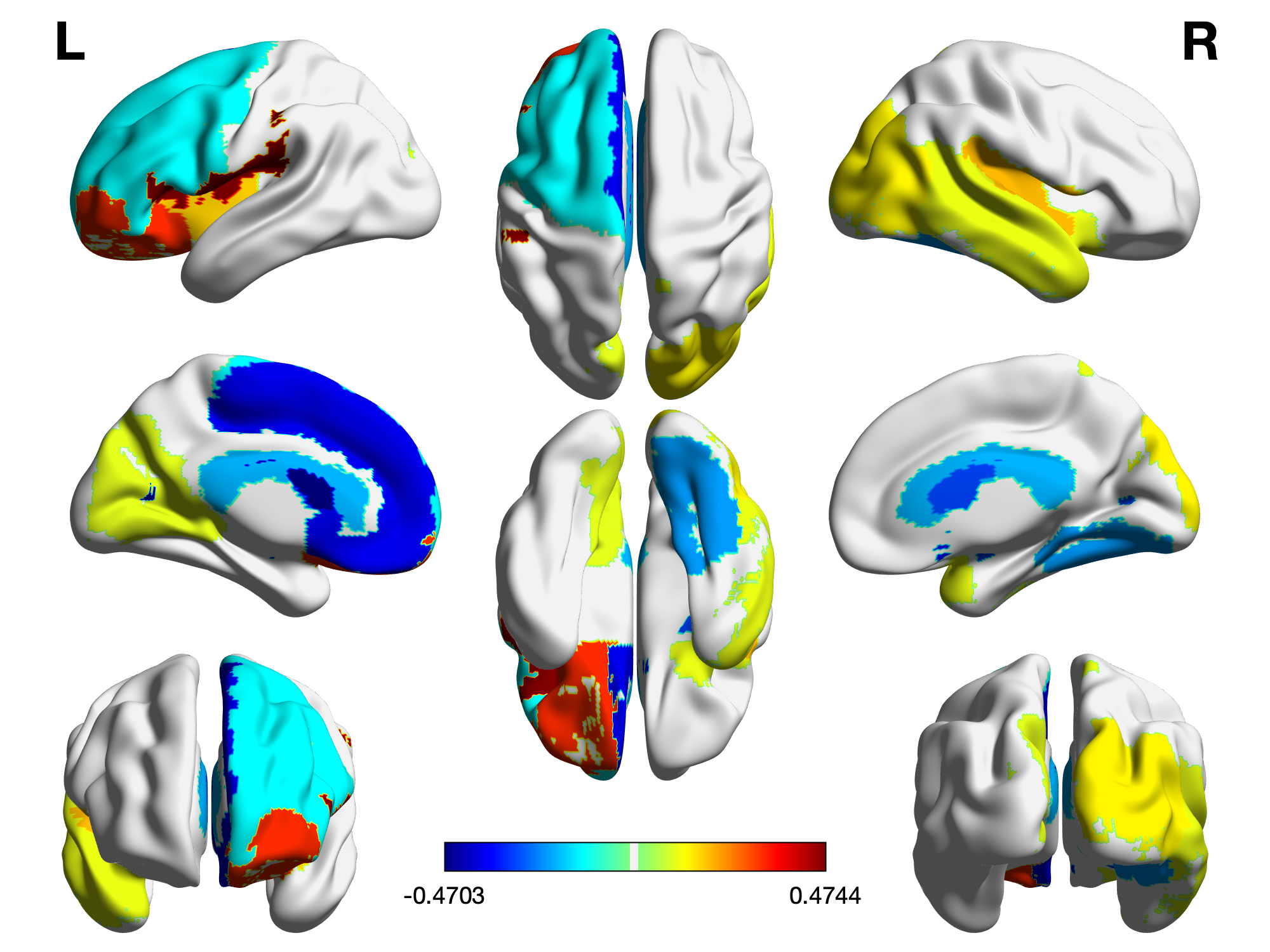}}
        \subfloat[\label{fig:adni_brain_map_C2}C2]{\includegraphics[width=0.33\textwidth]{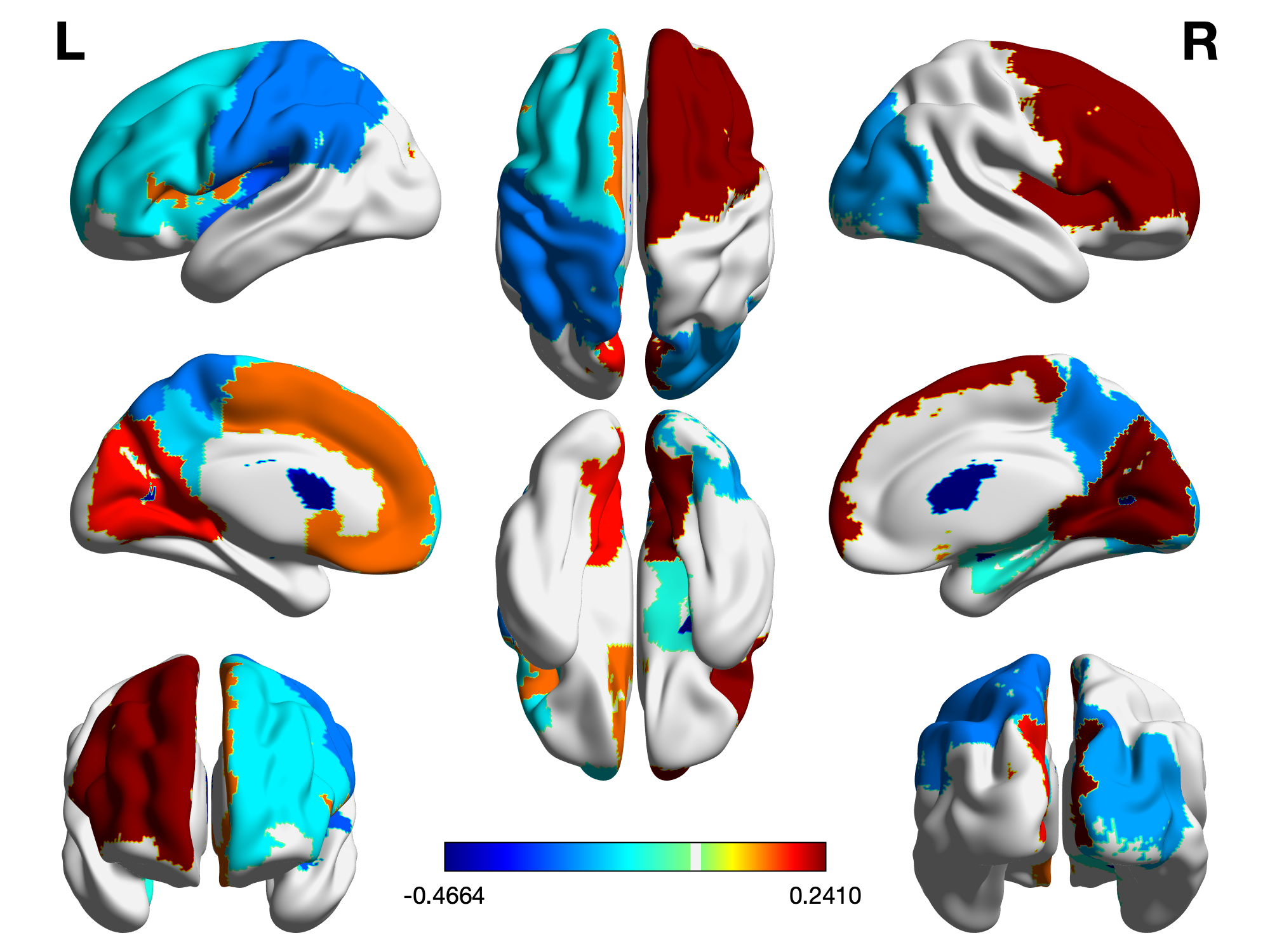}}
        \subfloat[\label{fig:adni_brain_map_C3}C3]{\includegraphics[width=0.33\textwidth]{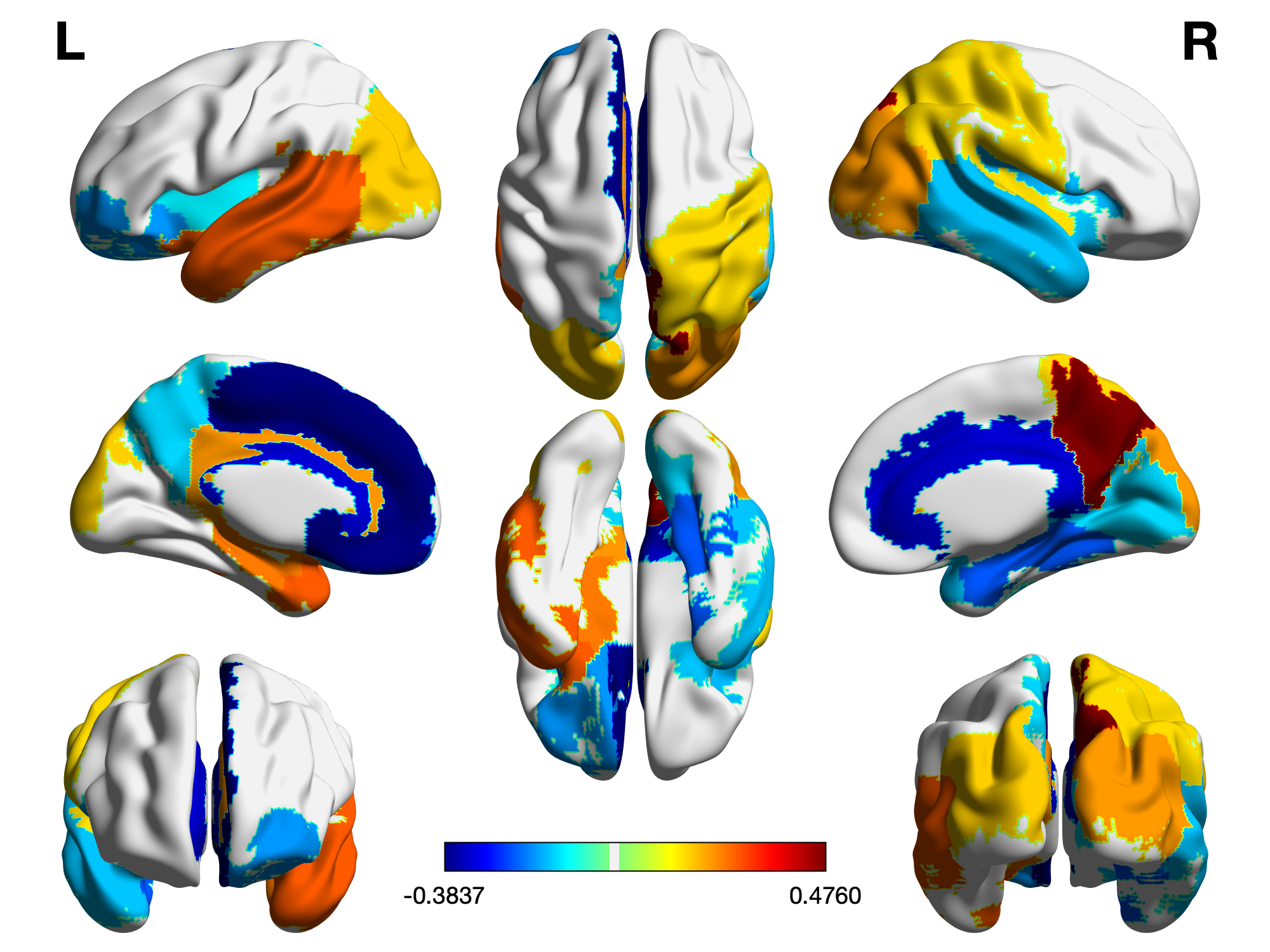}}
    \end{center}
    \caption{\label{fig:adni_brain}The brain map of the three identified components with a significant indirect effect.}
\end{figure}

Table~\ref{tabel:adni_brain} presents the top-loaded brain regions in the identified components. Figure~\ref{fig:adni_brain} shows these regions in a brain map. 
Some key regions include the frontal opercular, frontal medial, frontal inferior, frontal insula, frontal lateral, temporal lateral, temporal inferior, parietal lateral, parietal medial, occipital medial, limbic medial temporal, limbic cingulate, and ventricle. These regions cover cortical/subcortical areas including the pre/postcentral gyrus, cingulate gyrus, lingual gyrus, cuneus, inferior frontal gyrus, anterior/posterior insula, entorhinal cortex, parahippocampal gyrus, and hippocampus, which were all previously identified as marker regions with more severe atrophy in AD patients and associations with memory deficits were reported~\citep{nadel2011update,sadigh2014different,pini2016brain,parker2018cortical,jacobs2018cerebellum,schultz2018widespread}. The entorhinal and hippocampal atrophy are two well established and validated AD markers repeatedly reported in the existing literature. The entorhinal cortex was found to be affected by tau propagation first~\citep{braak1991neuropathological}. As a major component of the medial temporal lobe, the hippocampus involves in functions including response inhibition, episodic memory, and spatial cognition~\citep{jack2011steps}. Because of the sharp contrast between the CSF in the ventricles and surrounding tissues in brain structural images, volumetric measurement of the ventricles is robust to automatic segmentation. Enlargement in the ventricles is thus often considered a measurement of hemispheric atrophy rates and is consistently reported as an AD marker~\citep{nestor2008ventricular,kruthika2019multistage}.

In summary, the top-loaded features in the identified protein and brain components, as well as the direction of the effects, are in line with existing knowledge about AD.

\begin{table}
    \caption{\label{table:adni_coef}The estimates, standard error (SE), and $95\%$ bootstrap confidence interval (CI) from $1000$ samples of the model coefficients ($\alpha$ and $\beta$), indirect effect (IE), and direct effect (DE) of the three identified components.}
    \begin{center}
        \resizebox{\textwidth}{!}{
        \begin{tabular}{l r c c c r c c c r c c}
            \hline
            & \multicolumn{3}{c}{\textbf{C1}} && \multicolumn{3}{c}{\textbf{C2}} && \multicolumn{3}{c}{\textbf{C3}} \\
            \cline{2-4}\cline{6-8}\cline{10-12}
            & \multicolumn{1}{c}{Estimate} & \multicolumn{1}{c}{SE} & \multicolumn{1}{c}{$95\%$ CI} && \multicolumn{1}{c}{Estimate} & \multicolumn{1}{c}{SE} & \multicolumn{1}{c}{$95\%$ CI} && \multicolumn{1}{c}{Estimate} & \multicolumn{1}{c}{SE} & \multicolumn{1}{c}{$95\%$ CI} \\
            \hline
            $\alpha$ & $1.218$ & $0.070$ & $(1.083, 1.352)$ && $-1.052$ & $0.052$ & $(-1.156, -0.957)$ && $-1.110$ & $0.057$ & $(-1.224, -0.998)$ \\
            $\beta$ & $-0.807$ & $0.163$ & $(-1.134, -0.497)$ && $0.569$ & $0.167$ & $(0.247, 0.894)$ && $0.946$ & $0.200$ & $(0.569, 1.340)$ \\
            IE & $-0.984$ & $0.213$ & $(-1.416, -0.588)$ && $-0.600$ & $0.184$ & $(-0.971, -0.259)$ && $-1.051$ & $0.231$ & $(-1.524, -0.622)$ \\
            DE & $1.878$ & $0.231$ & $(1.456, 2.345)$ && $0.844$ & $0.209$ & $(0.438, 1.253)$ && $1.245$ & $0.292$ & $(0.682, 1.818)$ \\
            \hline
        \end{tabular}
        }
    \end{center}
\end{table}
\begin{table}
    \caption{\label{table:adni_protein}Proteins with top loading magnitude of the three components with a significant indirect effect. (Direction: protein regulation direction in MCI/AD in existing literature.)}
    \begin{center}
        \resizebox{\textwidth}{!}{
        \begin{tabular}{l L{6cm} r l c c L{6cm} r l c}
            \hline
            & \multicolumn{4}{c}{Positive Loading} && \multicolumn{4}{c}{Negative Loading} \\
            \cline{2-5}\cline{7-10}
            & \multicolumn{1}{c}{Protein} & \multicolumn{1}{c}{Loading} & \multicolumn{1}{c}{Gene} & \multicolumn{1}{c}{Direction} && \multicolumn{1}{c}{Protein} & \multicolumn{1}{c}{Loading} & \multicolumn{1}{c}{Gene} & \multicolumn{1}{c}{Direction} \\
            \hline
            & Neuronal pentraxin-2 & $0.358$ & NPTX2 & $\downarrow$ && ProSAAS & $-0.297$ & PCSK1N & $\Updownarrow$ \\
            & Neuronal cell adhesion molecule & $0.358$ & NRCAM & $\Updownarrow$ && Beta-2-microglobulin & $-0.284$ & B2M & $\Updownarrow$ \\
            & Neurexin-1 & $0.341$ & NRXN1 & $\downarrow$ && Chitinase-3-like protein 1 & $-0.268$ & CH3L1 & $\uparrow$ \\
            & Neurosecretory protein VGF & $0.318$ & VGF & $\downarrow$ && Kallikrein-6 & $-0.205$ & KLK6 & $\uparrow$ \\
            & Prostaglandin-H2 D-isomerase & $0.235$ & PTGDS & $\downarrow$ \\
            & Alpha-1B-glycoprotein & $0.165$ & A1BG & $\Updownarrow$ \\
            \multirow{-7}{*}{\textbf{C1}} & Calsyntenin-3 & $0.152$ & CSTN3 & $\Updownarrow$ \\
            \hline
            & Kallikrein-6 & $0.533$ & KLK6 & $\uparrow$ && Brain acid soluble protein 1 & $-0.287$ & BASP1 & $\Updownarrow$ \\
            & Clusterin & $0.405$ & CLUS & $\Updownarrow$ && Alpha-1B-glycoprotein & $-0.277$ & A1BG & $\Updownarrow$ \\
            & Insulin-like growth factor-binding protein 2 & $0.225$ & IGFBP2 & $\Updownarrow$ && Apolipoprotein D & $-0.229$ & APOD & $\Updownarrow$ \\
            & & & & && Apolipoprotein E & $-0.229$ & APOE & $\Updownarrow$ \\
            \multirow{-6}{*}{\textbf{C2}} & & & & && Glial fibrillary acidic protein & $-0.189$ & GFAP & $\uparrow$\\
            \hline
            & ProSAAS & $0.420$ & PCSK1N & $\Updownarrow$ && Alpha-1B-glycoprotein & $-0.347$ & A1BG & $\Updownarrow$ \\
            & Prostaglandin-H2 D-isomerase & $0.293$ & PTGDS & $\downarrow$ && Beta-2-microglobulin & $-0.212$ & B2M & $\Updownarrow$ \\
            & Monocyte differentiation antigen CD14 & $0.243$ & CD14 & $\uparrow$ && Mimecan & $-0.197$ & OGN & $\Updownarrow$ \\
            & Compliment factor B & $0.243$ & CFB & $\uparrow$ && Neurosecretory protein VGF & $-0.185$ & VGF & $\downarrow$ \\
            & Neuronal pentraxin-1 & $0.242$ & NPTX1 & $\downarrow$ && Ectonucleotide pyrophosphatase/phosphodilesterase family member 2 & $-0.181$ & ENPP2 & $\uparrow$ \\
            & Calsyntenin-3 & $0.196$ & CSTN3 & $\Updownarrow$ && Chromogranin-A & $-0.178$ & CHGA & $\Updownarrow$ \\
            & Complement C4-A & $0.183$ & C4A & $\uparrow$ && Cystatin-C & $-0.176$ & CST3 & $\Updownarrow$ \\
            \multirow{-11}{*}{\textbf{C3}} & & & & && Gamma-enolase & $-0.176$ & ENO2 & $\Updownarrow$ \\
            \hline
        \end{tabular}
        }
    \end{center}
    \vspace{-1ex}
    {\raggedright {\scriptsize{$\uparrow$/$\downarrow$: consistently upregulated/downregulated in MCI/AD; $\Updownarrow$: inconsistent reports.}}}
\end{table}
\begin{table}
    \caption{\label{tabel:adni_brain}Brain regions with top loading magnitude of the three identified components with a significant indirect effect. (L: Left; R: Right; WM: white matter; GM: gray matter.)}
    \begin{center}
        \resizebox{\textwidth}{!}{
        \begin{tabular}{l l l r c l l r}
            \hline
            & \multicolumn{3}{c}{Positive Loading} && \multicolumn{3}{c}{Negative Loading} \\
            \cline{2-4}\cline{6-8}
            & \multicolumn{1}{c}{Region} & \multicolumn{1}{c}{Module} & \multicolumn{1}{c}{Loading} && \multicolumn{1}{c}{Region} & \multicolumn{1}{c}{Module} & \multicolumn{1}{c}{Loading} \\
            \hline
            & Frontal opercular (L) & Frontal & $0.474$ && Ventricle (L) & Ventricle & $-0.470$ \\
            & Frontal inferior (L) & Frontal & $0.319$ && Frontal medial (L) & Frontal & $-0.371$ \\
            & Temporal supratemporal (R) & Temporal & $0.189$ && Ventricle (R) & Ventricle & $-0.293$ \\
            & Frontal insular (L) & Frontal & $0.168$ && Cerebellum (R) & Cerebellum & $-0.214$ \\
            & Occipital lateral (R) & Occipital & $0.135$ && Corpus callosum & Corpus callosum & $-0.194$ \\
            & Temporal lateral (R) & Temporal & $0.107$ && Frontal lateral (L) & Frontal & $-0.125$ \\
            \multirow{-7}{*}{\textbf{C1}} & Occipital medial (L) & Occipital & $0.101$ \\
            \hline
            & Occipital medial (R) & Occipital & $0.241$ && Ventricle (R) & Ventricle & $-0.466$ \\
            & Frontal insular (R) & Frontal & $0.234$ && Ventricle (L) & Ventricle & $-0.408$ \\
            & Frontal lateral (R) & Frontal & $0.233$ && Temporal supratemporal (L) & Temporal & $-0.277$ \\
            & Occipital medial (L) & Occipital & $0.179$ && Parietal lateral (L) & Parietal & $-0.242$ \\
            & Frontal medial (L) & Frontal & $0.135$ && Parietal medial (R) & Parietal & $-0.232$ \\
            & Frontal opercular (L) & Frontal & $0.135$ && Occipital lateral (R) & Occipital & $-0.205$ \\
            & & & && Occipital inferior (R) & Occipital & $-0.185$ \\
            & & & && Parietal medial (L) & Parietal & $-0.159$ \\
            & & & && Frontal insular (L) & Frontal & $-0.135$ \\
            & & & && Frontal lateral (L) & Frontal & $-0.135$ \\
            \multirow{-11}{*}{\textbf{C2}} & & & && Limbic medial temporal (R) & Limbic & $-0.101$ \\
            \hline
            & Parietal medial (R) & Parietal & $0.476$ && Frontal medial (L) & Frontal & $-0.384$ \\
            & Temporal lateral (L) & Temporal & $0.277$ && Limbic cingulate (R) & Limbic & $-0.322$ \\
            & Limbic medial temporal (L) & Limbic & $0.247$ && Temporal inferior (R) & Temporal & $-0.228$ \\
            & Limbic cingulate (L) & Limbic & $0.221$ && Frontal inferior (L) & Frontal & $-0.180$ \\
            & Occipital lateral (R) & Occipital & $0.218$ && Frontal opercular (R) & Frontal & $-0.156$ \\
            & Occipital lateral (L) & Occipital & $0.171$ && Temporal lateral (R) & Temporal & $-0.147$ \\
            & Parietal lateral (R) & Parietal & $0.157$ && Parietal medial (L) & Parietal & $-0.134$ \\
            & & & && Occipital medial (R) & Occipital & $-0.131$ \\
            \multirow{-9}{*}{\textbf{C3}} & & & && Frontal insular (L) & Frontal & $-0.115$ \\
            \hline
        \end{tabular}
        }
    \end{center}
\end{table}

\section{Discussion}
\label{sec:discussion}

In this study, a linear structural equation modeling framework is proposed for mediation analysis with multiple exposures and multiple mediators. The framework assumes that there exist underlying orthogonal mediation mechanisms on the outcome, thus named \textbf{P}rincipal \textbf{C}omponent \textbf{M}ediation \textbf{A}nalysis (PCMA). Under the normality assumption, a likelihood-based approach is proposed to simultaneously estimate the orthogonal projections and effect parameters. Under a low-dimensional scenario, the asymptotic distributions of the proposed estimators are derived. A bootstrap procedure is also introduced for finite sample inference. Simulation studies demonstrate the superior performance of the proposal compared to a competing approach. In the ADNI proteomic-imaging study, the proposed approach identifies protein--brain structure components that have a significant mediation effect on memory decline among MCI patients. Features in the components are consistent with existing knowledge about AD and suggest pathological paths of CSF protein deposition -- brain atrophy -- memory deficit.

The proposed framework assumes the existence of latent components that connect the exposures, mediators, and outcome. The identified components are linear combinations of the exposure/mediator features making the interpretation less feasible. Though an \textit{ad hoc} approach is suggested to sparsify the loading profiles, integrated approaches can be a future direction. This type of approach may also apply to the scenario of high-dimensional exposures and/or high-dimensional mediators. In the current study, asymptotic properties are derived under the low-dimensional scenario. Asymptotic theories under the high-dimensional setting are challenging without any constraint or regularization and thus are left to future research. The proposed estimators are likelihood-based estimators. In practice, when properly scaling the data and imposing unit variances ($\sigma=\tau=1$), it is equivalent to the least squares estimation. The consistency still holds but the estimators are more robust to non-Gaussian continuous distributions~\citep{charnes1976equivalence,white1980heteroskedasticity}. For other types of data outcome, such as a binary outcome, extensions to the generalized SEMs are feasible but require further investigation. The proposed framework assumes no interaction between the exposures and mediators. With multiple exposures and mediators, an extension of including interactions is not straightforward and is considered a future direction.

\section*{Acknowledgments}

\bibliographystyle{apalike}
\bibliography{Bibliography}

\end{document}